\documentclass[twocolumn,prlprintnumbers,showpacs,prl,floatfix,amsmath,amssymb,notitlepage,superscriptaddress]{revtex4-1}

\usepackage{epsfig}
\usepackage{subfigure}
\usepackage{amsmath}
\usepackage{color}
\usepackage{amssymb}
\usepackage{setspace}
\usepackage{graphicx}% Include figure files
\usepackage{dcolumn}% Align table columns on decimal point
\usepackage{bm}% bold math
\usepackage{times}
\usepackage{enumerate}
\usepackage{footnote}
\input epsf

\graphicspath{{figures/}}

\renewenvironment{widetext@grid}{%
  \par\ignorespaces
  \setbox\widetext@top\vbox{%
   \vskip15\p@
   \hb@xt@\hsize{%
    \leaders\hrule\hfil
    \vrule\@height6\p@
   }%
   \vskip6\p@
  }%
  \setbox\widetext@bot\hb@xt@\hsize{%
    \vrule\@depth6\p@
    \leaders\hrule\hfil
  }%
  \onecolumngrid
  \let\set@footnotewidth\set@footnotewidth@ii
}{%
  \par
  \twocolumngrid\global\@ignoretrue
  \@endpetrue
}
\makeatother

\begin{document}

\author{Dane Taylor}
\email{dane.r.taylor@gmail.com}
\affiliation{Carolina Center for Interdisciplinary Applied Mathematics, Department of Mathematics, University of North Carolina, Chapel Hill, NC 27599, USA}

\author{Saray Shai}
\affiliation{Carolina Center for Interdisciplinary Applied Mathematics, Department of Mathematics, University of North Carolina, Chapel Hill, NC 27599, USA}

\author{Natalie Stanley}
\affiliation{Carolina Center for Interdisciplinary Applied Mathematics, Department of Mathematics, University of North Carolina, Chapel Hill, NC 27599, USA}
\affiliation{Curriculum in Bioinformatics and Computational Biology, University of North Carolina, Chapel Hill, NC 27599, USA}

\author{Peter J. Mucha}
\affiliation{Carolina Center for Interdisciplinary Applied Mathematics, Department of Mathematics, University of North Carolina, Chapel Hill, NC 27599, USA}

\title{Enhanced detectability of community structure in {multilayer} networks through layer aggregation}

\begin{abstract}
Many systems are naturally represented by a multilayer network in which edges exist in multiple layers that encode different, but potentially related, types of interactions, and it is important to understand limitations on the detectability of community structure in these networks. Using random matrix theory, we analyze detectability limitations for {multilayer (specifically, multiplex) stochastic block models (SBMs) in which $L$ layers are derived from a common SBM. We study the effect of layer aggregation on detectability for several aggregation methods,} including summation of the layers' adjacency matrices for which we show the detectability limit vanishes as $\mathcal{O}(L^{-1/2})$ with increasing number of layers, $L$. Importantly, we find a similar scaling behavior when the summation is thresholded at an optimal value, {providing insight into} the common---but not well understood---practice of thresholding {pairwise-interaction} data to obtain sparse network representations. 

\end{abstract}


\maketitle

The analysis of complex networks \cite{Newman2003} has far-reaching applications ranging from social systems \cite{Moody2004} to the brain \cite{Bassett2011}. Often, a natural representation is that of a multilayer network (see reviews \cite{Boccaletti2014,Kivela2014}), whereby network layers encode different classes of interactions, such as categorical social ties~\cite{Krackhardt1987}, types of critical infrastructure \cite{Haimes}, or a network at different instances in time \cite{Holme2012}. In principle, the multilayer framework offers a more comprehensive representation of a data set or system, as compared to an aggregation of network layers that produces a simplified model but does so at the cost of information loss. For example, neglecting the layered structure can lead to severe and unintended consequences regarding structure \cite{Mucha2010} and dynamics \cite{Sole_2013,Sanchez_2014,Brummitt2012}, which can fundamentally differ {between} single-layer {and multilayer} networks \cite{Radicchi2015,Bashan2013}.

However, layer aggregation also implements an information processing that can yield beneficial results. Network layers are often correlated with one another {and can encode redundant information \cite{Menichetti2014}.} In some cases a multilayer representation is an over-modeling, which can negatively impact the computational and memory requirements for storage and analysis. In such situations, it is beneficial to seek a more concise representation in which certain layers are aggregated \cite{Domenico2015,Stanley2015}. Identifying sets of repetitive layers amounts to a clustering problem, and it is closely related to the topic of clustering networks in an ensemble of networks \cite{Onnela2012,Stanley2015}. Much remains to be studied regarding \emph{when} layer aggregation is appropriate and \emph{how} it should be implemented. 
%Importantly, we expect answers to such questions to depend on the application at hand.

We study here the effect of layer aggregation on community structure in multilayer networks in which each layer is drawn from a common stochastic block model (SBM). SBMs are a paradigmatic model \cite{Lancichinetti_2009} for complex structure in networks and are particularly useful for studying limitations on detectability---that is, if the community structure is too weak, it cannot be found upon inspection of the network \cite{Hu2012,Lance2012,Reichardt2008,Decelle_2011,Nadakuditi_2012,Abbe2014}. Recently, the detectability limit has been explored for networks with degree heterogeneity \cite{Radicchi_2013} and hierarchical structure {\cite{Peixoto2013,Sarkar2013}}, for temporal networks \cite{Ghasemian2015}, and for the detection of communities using multi-resolution methods \cite{Kawamoto_2015}. Despite growing interest in multilayer SBMs \cite{Peixoto2015_a,Paul2015,Barbillon2015,Valles2014,Han2015} (which we note, focus on \emph{multiplex} networks in which nodes are identical in every layer and edges are restricted to connecting nodes in the same layer \cite{Boccaletti2014,Kivela2014}), the effect of layer aggregation on detectability limitations has yet to be explored outside the infinite layer limit \cite{Han2015}. 

To this end, we study detectability limitations for multilayer SBMs {with layers following from identical SBM parameters} and find that the method of aggregation significantly influences detectability. When the aggregate network corresponds to the summation of the adjacency matrices encoding the network layers, aggregation always improves detectability. In particular, the detectability limit vanishes with increasing number of layers, $L$, and decays as $\mathcal{O}(L^{-1/2})$. Because the summation of $L$ adjacency matrices can often yield a weighted and dense network---which increases the complexity of community detection \cite{Aicher2015}---we also study binary adjacency matrices obtained by thresholding this summation at some value $\tilde{L}$. We find that the detectability limit is very sensitive to the choice of $\tilde{L}$; however, we also find that there exist thresholds (e.g., mean edge probability for homogeneous communities) that are optimal in that the detectability limit also decays as $\mathcal{O}(L^{-1/2})$. 
These results {provide insight into} the use of thresholding pairwise-interaction data so as to produce sparse networks---a practice that is commonplace but for which the effects are not well understood.

We begin by describing the multilayer SBM. We consider $N$ nodes divided into $K$ communities, and we denote by $c_i\in\{1,\dots,K\}$ the community index for each node $i\in\{1,\dots,N\}$. The multiplex network is defined by $L$ layers encoded by a set of adjacency matrices, $\{{\bf A}^{(l)}\}$, where $A^{(l)}_{ij}=1$ if $(i,j)$ is an edge in layer $l$ and $A^{(l)}_{ij}=0$ otherwise. The probability of edge $(i,j)$ in layer $l$ is given by $\Pi_{c_{i}c_{j}} \in[0,1]$, where ${\bf \Pi}$ is a $K\times K$ matrix. 

The detectability of community structure relates to the ability to recover the nodes' community labels $\{c_i\}$. 
To connect with previous research {\cite{Reichardt2008,Decelle_2011,Nadakuditi_2012,Abbe2014},} we focus on the case of $K=2$ communities of equal size with edge probabilities $\Pi_{11} =\Pi_{22}=p_{in}$ and $\Pi_{12} =\Pi_{21}=p_{out}$. 
{Below, we will simultaneously refer to these respective probabilities as $p_{in,out}$.}
We assume $p_{in}\ge p_{out}$ to study ``assortative'' communities in which there is a prevalence of edges between nodes in the same community \cite{Rombach2014}. 

It has been shown for the large network $N\to\infty$ limit that there exists a detectability limit characterized {\cite{Decelle_2011,Nadakuditi_2012}} by the solution curve $(\Delta^*,\rho)$ to 
\begin{equation}
N \Delta = \sqrt{4 N \rho},\label{eq:detectability}
\end{equation} 
where $\Delta=p_{in}-p_{out}$ is the difference in probability and $\rho=(p_{in}+{p_{out}})/2$ is the mean edge probability. For given $\rho$, the communities are detectable only when the presence of community structure is sufficiently strong, {i.e.,} $\Delta>\Delta^*$. 
Equation~\eqref{eq:detectability} describes a phase transition that has been obtained via complementary analyses---Bayesian inference \cite{Decelle_2011} and random matrix theory \cite{Nadakuditi_2012}---and represents a critical point that is independent of the community detection method (see \cite{Decelle_2011} and footnote 11 in \cite{Nadakuditi_2012}). We further note that Eq.~\eqref{eq:detectability} was derived for sparse networks [i.e., constant $\rho N$ so that $\rho=\mathcal{O}(N^{-1})$]. Here, we must consider the full range of densities, $\rho\in[0,1]$, to allow for aggregated networks that are potentially dense [i.e., $\rho=\mathcal{O}( 1)$ as $N\to\infty$].

In this Letter, we study the behavior of $\Delta^*$ for two methods of aggregating layers. We define the \emph{summation} network corresponding to the weighted adjacency matrix ${\bf \overline{A}} = \sum_l {\bf A}^{(l)}$ {as well as} a family of \emph{thresholded} networks {with} unweighted adjacency matrices $\{{\bf\hat{A}}^{(\tilde{L})}\}$ that are obtained by applying a threshold $\tilde{L}\in\{1,\dots,L\}$ to the entries of ${\bf \overline{A}}$. {Specifically, we define} $\hat{A}^{(\tilde{L})}_{ij} = 1$ if $\overline{A}_{ij}\ge \tilde{L}$ and $\hat{A}^{(\tilde{L})}_{ij}=0$ otherwise. Of particular interest are the limiting cases $\tilde{L}=L$ and $\tilde{L}=1$, which respectively correspond to applying logical AND and OR operations to the {original multiplex data} $\{A_{ij}^{(l)}\}$ for fixed $(i,j)$. We refer to these thresholded networks as the {AND} and {OR} networks, respectively.

We study the detectability limit for the layer-aggregated networks using random matrix theory \cite{Benaych_2011,Nadakuditi_2013}. This approach is particularly suited for detectability {analysis} since community labels $\{c_i\}$ can be identified using spectral partitioning and phase transitions {\cite{Nadakuditi_2012,Peixoto2013,Sarkar2013}} in detectability correspond to the disappearance of gaps between isolated eigenvalues (whose corresponding eigenvectors reflect community structure) and bulk eigenvalues [which arise due to stochasticity and whose $N\to\infty$ limiting distribution is given by a spectral density $P(\lambda)$]. We develop theory based on the modularity matrix $\overline{B}_{ij}=\overline{A}_{ij}-\rho L$ \cite{Newman_2004}. Note that we do not use the configuration model as the null model. Instead, since all nodes are identical under the SBM, the appropriate null model is Erd\H{o}s-R\'enyi {with repeated edges allowed} in which the expected number of edges between any pair of nodes is $\rho L$.  

We first study $\Delta^*$ for the summation network. We analyze the distribution of real eigenvalues $\{\lambda_i\}$ of ${\bf \overline{B}}$ (in descending order) 
using methodology developed in \cite{Nadakuditi_2012,Benaych_2011}; we extend this work to networks that are multiplex and possibly dense. We outline our results here and provide further details in the Supplemental Material. We begin by describing the statistical properties of entries $\{\overline{A}_{ij}\}$, which are independent random variables following a binomial distribution $P\left(\overline{A}_{ij} =a\right) = f(a;L,\Pi_{c_ic_j})$, {where}
\begin{equation}
 f(a;L,p)= \left(\footnotesize{\begin{array}{c}L\\ a\end{array}}\right) {p^a} (1-p)^{L-a}\label{eq:Entry_dist}
\end{equation}
has mean $Lp$ and variance $Lp(1-p)$. Provided that there is sufficiently large variance in the edge probabilities (i.e., $NL\rho(1-\rho)\gg1$), we find that the limiting $N\to\infty$ distribution of bulk eigenvalues for ${\bf \overline{B}}$ is given by a semi-circle distribution,
\begin{equation}
P({\lambda})= \frac{{\sqrt{\lambda_2^2 -\lambda^2}}}{\pi \lambda_2^2/2} \label{eq:density}
\end{equation} 
for $|\lambda|<\lambda_2$ and $P(\lambda)=0$ otherwise, where
\begin{align}
{\lambda}_2 = \sqrt{4NL[\rho(1-\rho) {-\Delta^2/4}]  \label{eq:Max_bulk}} 
\end{align}
is the upper bound on the support of this spectral density and is the limiting $N\to\infty$ value of the second-largest eigenvalue. The largest eigenvalue {of ${\bf \overline{B}}$ in the $N\to\infty$ limit} is an isolated eigenvalue
\begin{align}
{\lambda}_1 = NL\Delta/2 + 2[\rho(1-\rho) {-\Delta^2/4}] /\Delta \label{eq:Max_eval}.
\end{align}
As we shall show, $\Delta^*\to0$ as $N$ increases, and therefore the $\Delta^2/4$ terms in Eq.~\eqref{eq:Max_bulk} and \eqref{eq:Max_eval} are negligible near the detectability limit (i.e., $\Delta\approx\Delta^*)$.
The eigenvector $\bf{v}$ corresponding to $\lambda_1$ gives the spectral bipartition---the inferred community label of node $i$ is determined by the sign of $v_i$---and provided that the largest eigenvalue corresponds to this isolated eigenvalue, $\lambda_1$, the eigenvector entries $\{v_i\}$ are correlated with the community labels $\{c_i\}$. 
To obtain the detectability limit, we set $\lambda_1=\lambda_2$, neglect the $\Delta^2/4$ terms and simplify, yielding a modified detectability equation 
\begin{equation}
NL \Delta = \sqrt{4 NL \rho(1-\rho)}.\label{eq:Delta}
\end{equation} 
Note that Eq.~\eqref{eq:Delta} recovers Eq.~\eqref{eq:detectability} when $L=1$ and $\rho\to0$ [i.e., for sparse networks, $\rho(1-\rho) \approx \rho$].  Defining $p_{in}^*=\rho+\Delta^*/2$ and $p_{out}^*=\rho-\Delta^*/2$, we find for fixed $\rho$ and increasing $N$ and/or $L$ that {$p_{{in,out}}^*\to\rho$} and $\Delta^*\to0$, decaying as $\mathcal{O}(1/\sqrt{NL})$.

We now study $ \Delta^*$ for the thresholded networks, which correspond to single-layer SBMs in which the community labels $\{c_j\}$ are identical to those of the multilayer SBM, but there are new \emph{effective} block edge probabilities 
\begin{equation}
\hat{\Pi}_{nm}^{(\tilde{L})}=1-F(\tilde{L}-1;L,\Pi_{nm}),\label{eq:pi}
\end{equation}
where $F(a;L,p)$ is the cumulative distribution function for the binomial distribution $f(a;L,p)$. The effective probabilities for the AND and OR networks are $\hat{\Pi}_{nm}^{(L)}= {(\Pi_{nm})^L }$ and $\hat{\Pi}_{nm}^{(1)}= 1-(1-\Pi_{nm})^L $, respectively. For the two-community SBM, the effective probabilities are $\hat{p}_{{in,out}}^{(\tilde{L})} = 1-F(\tilde{L}-1;L,p_{{in,out}})$, $\hat{\Delta}^{(\tilde{L})} = \hat{p}_{in}^{(\tilde{L})}-\hat{p}_{out}^{(\tilde{L})}$, and $\hat{\rho}^{(\tilde{L})}=(\hat{p}_{in}^{(\tilde{L})}+\hat{p}_{out}^{(\tilde{L})})/2$. The modularity matrices for the thresholded networks become $\hat{B}_{ij}^{(\tilde{L})}=\hat{A}_{ij}^{(\tilde{L})}-\hat{\rho}^{(\tilde{L})} $. We identify the detectability limit by substituting $\hat{\Delta}^{(\tilde{L})}\mapsto \Delta$ and $\hat{\rho}^{(\tilde{L})}\mapsto\rho$ into Eq.~\eqref{eq:Delta} (with $L=1$) and numerically finding a solution $(\Delta^*,\rho)$ using a root-finding algorithm. Note that the detectability equation holds for the effective probabilities, $N\hat{\Delta}^{(\tilde{L})} = \sqrt{4N\hat{\rho}^{(\tilde{L})}(1-\hat{\rho}^{(\tilde{L})})}$, and not the single-layer probabilities, $N{\Delta} \not= \sqrt{4N{\rho}(1-\rho)}$.

\begin{figure}[t]
\centering
\epsfig{file = 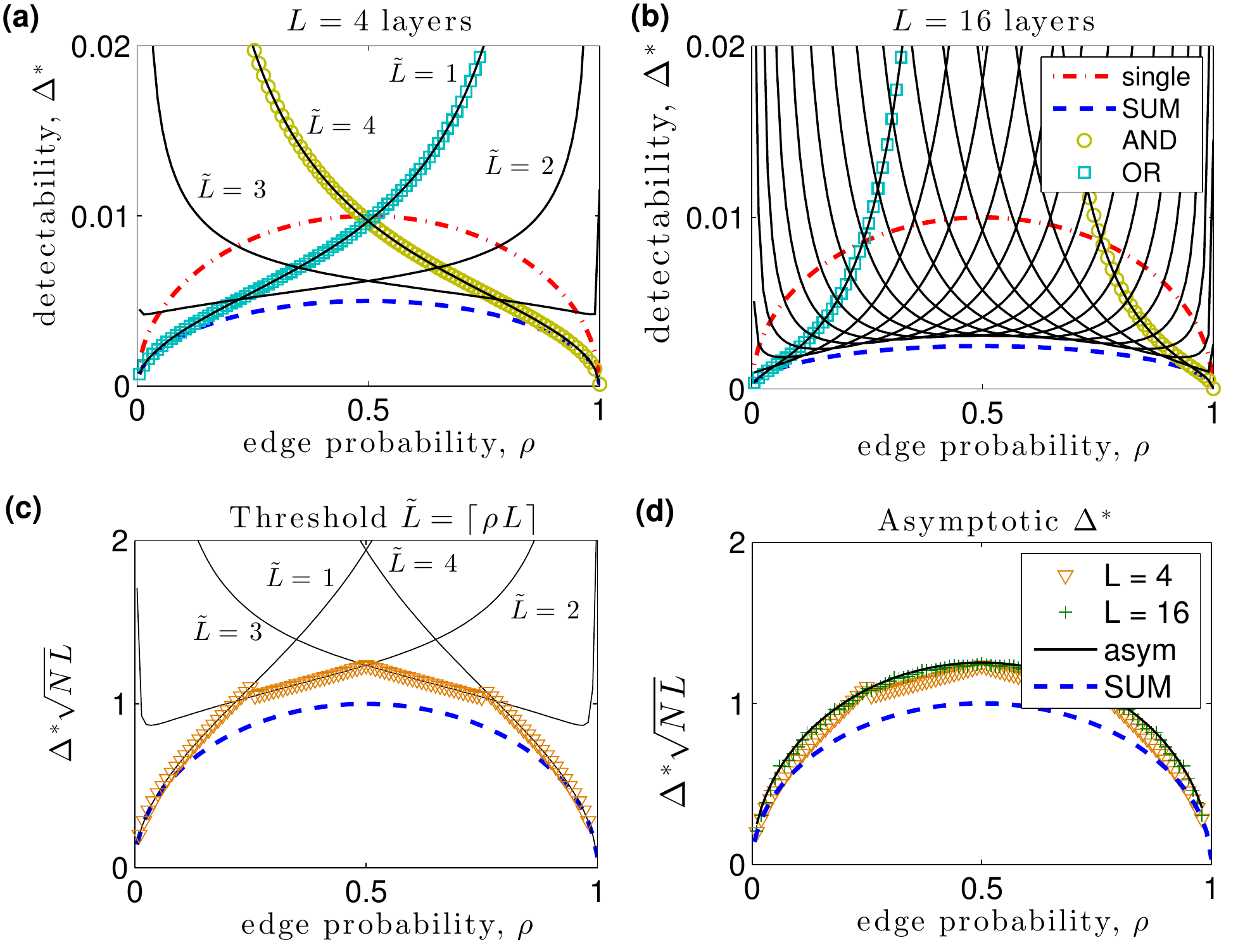, clip =,width=1\linewidth } 
\caption{(Color online) {\it Layer aggregation enhances the detectability of community structure.}
(a)--(b) We plot {the detectability limit} $\Delta^*$ versus mean edge probability $\rho$ for a single network layer (red dot-dashed curves), {the aggregate network obtained by summation} (blue dashed curves), and {aggregate networks obtained by thresholding this summation at $\tilde{L}\in\{1,2,3,4\}$} (solid curves). Gold circles and cyan squares highlight $\tilde{L}=L$ and $\tilde{L}=1$, which we refer to as AND and OR networks, respectively.
Results are shown for {$N=10^4$} nodes with (a) $L=4$ and (b)~$L=16$ layers. 
(c) {For $L=4$,} we show $\Delta^*$ versus $\rho$ for the {optimal threshold} $\tilde{L}=\lceil \rho L\rceil$ (orange triangles), which lies on the solution curves for {$\tilde{L}\in\{1,\dots,L\}$} (solid curves). 
(d) We show $\Delta^*$ for $\tilde{L}=\lceil \rho L\rceil$ with $L\in\{4,16\}$. These piecewise-continuous solutions collapse onto the asymptotic solution $\Delta^*_{(asym)}$ (black curve) as $L$ increases. In panels (c)--(d), we additionally plot $\Delta^*$ for the summation network (blue dashed curves).
} 
\label{fig:Multiple_thresholds}
\end{figure}

In Figs.~\ref{fig:Multiple_thresholds}(a)--(b), we show $\Delta^*$ versus the mean edge probability $\rho$ for the different aggregation methods: 
{(i)}~a single layer (red dot-dashed curves), which is identical in panels (a) and (b);
{(ii)}~the summation network (blue dashed curves), for which the curve in (b) corresponds to the curve in panel (a) rescaled by a factor of $1/2$;
{and (iii)}~thresholded networks (solid curves), which shift left-to-right with increasing $\tilde{L}$.
This is evident by comparing $\Delta^*$ for the AND ($\tilde{L}=L$, gold circles) and OR ($\tilde{L}=1$, cyan squares) networks. We find when $\rho$ is large that the AND (OR) network has a relatively small (large) detectability limit; in contrast, when $\rho$ is small the AND (OR) network has a relatively large (small) detectability limit. In other words, aggregating layers using the AND (OR) operation is beneficial for dense (sparse) networks.

It is interesting to ask if there are choices of $\rho$ and $\tilde{L}$ for which the detectability limit vanishes as $\mathcal{O}(L^{-1/2})$ with increasing $L$---that is, a behavior similar to that of the summation network. To this end, we study the threshold $\tilde{L}=\lceil \rho L\rceil$, which we {numerically observe} to be the {best $\tilde{L}$} for most values of $\rho$. This choice is also convenient as it only requires knowledge of the mean edge probability, $\rho$, which is easy to obtain in practice. In Fig.~\ref{fig:Multiple_thresholds}(c), we plot $\Delta^*$ versus $\rho$ for $L=4$ and $\tilde{L}=\lceil \rho L\rceil$ (orange triangles), which lies along the solution curves for $\tilde{L}\in\{1,\dots,L\}$ (solid curves). 
In Fig.~\ref{fig:Multiple_thresholds}(d), we plot $\Delta^*$ for threshold $\tilde{L}=\lceil \rho L\rceil$ with $L=4$ (orange triangles) and $L=64$ (green crosses). These curves align due to the rescaling of the vertical axis by $\sqrt{NL}$. In fact, we find in the large $L$ limit that these solutions $\Delta^*$ collapse onto a single 
{curve $(\Delta^*_{(asym)},\rho)$ that solves
\begin{align}
NL\Delta  &=  \sqrt{2\pi NL \rho(1-\rho)} , \label{eq:ASYMPOTIC}
\end{align}
which we plot by the black line in Fig.~\ref{fig:Multiple_thresholds}(d). To obtain Eq.~\eqref{eq:ASYMPOTIC}, we use the central limit theorem \cite{CLT} to approximate $\hat{p}_{{in,out}}^{(\lceil \rho L\rceil)}\approx \hat{p}_{{in,out}}^{(asym)} = 1-G\left(L\rho;Lp_{in,out},Lp_{in,out}\left(1-p_{{in,out}}\right)\right)$,
%
%for large $L$, where
%\begin{align}
%\hat{p}_{{in,out}}^{(asym)} &\triangleq  1-G\left(L\rho;Lp_{in,out},Lp_{in,out}\left(1-p_{{in,out}}\right)\right) %\nonumber\\
%\end{align}
%and 
where $G\left(p;\mu,\sigma^2\right)=\frac{1}{2} + \frac{1}{2}\text{erf}\left(  {(p-\mu)}/{\sigma\sqrt{2}}\right)$ is the value of the cumulative distribution function of the normal distribution with mean $\mu$ and variance $\sigma^2$ evaluated at $p$. In particular, we approximate $\hat{\Delta}^{(\lceil \rho L\rceil)}\approx \hat{\Delta}^{(asym)}=\text{erf}\left(  \Delta\sqrt{L}/{\sqrt{8\rho(1-\rho)}}\right)$ and $\hat{\rho}^{(\lceil \rho L\rceil)}\approx\hat{\rho}^{(asym)}=1/2$. Equation ~\eqref{eq:ASYMPOTIC} is recovered after substituting $\hat{\Delta}^{(asym)}\mapsto \Delta$ and $\hat{\rho}^{(asym)}\mapsto\rho$ into Eq.~\eqref{eq:Delta} with $L=1$ and using the first-order expansion $\text{erf}^{-1}(N^{-1/2})\approx \sqrt{\pi/4N}$. Importantly, Eq.~\eqref{eq:ASYMPOTIC} implies that $\Delta^*$ decays as $\mathcal{O}(1/\sqrt{NL})$ for thresholded networks with $\tilde{L}=\lceil \rho L \rceil$. }

\begin{figure}[t]
\centering
\epsfig{file = 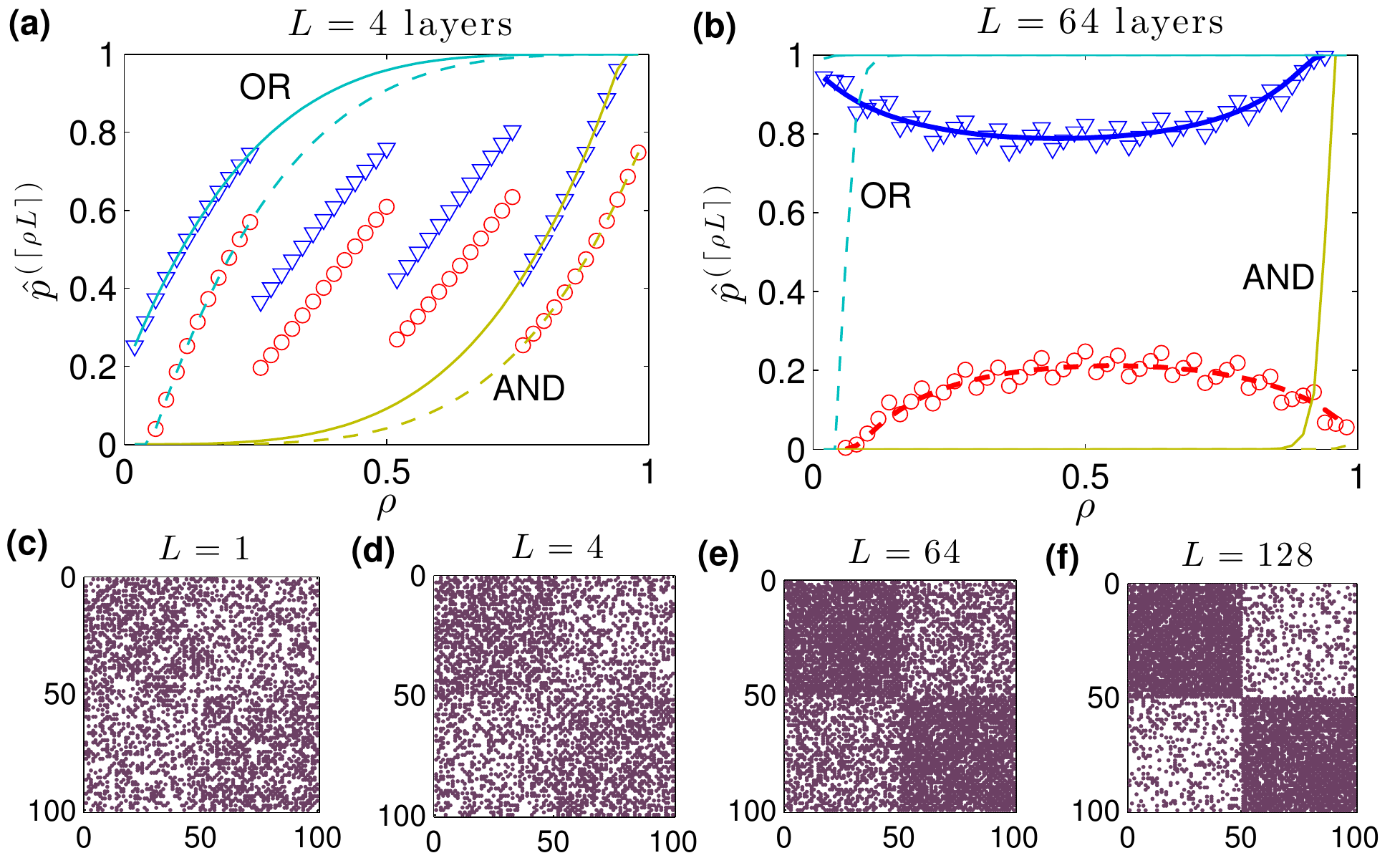, clip =,width=1\linewidth } 
\caption{(Color online) 
{\it Effective edge probabilities for layer aggregation at an optimal threshold.}
(a)--(b) {The summation and thresholding at $\tilde{L}=\lceil \rho L\rceil$ of $L$ adjacency matrices yields a new SBM with effective edge probabilities $\hat{p}_{in}^{(\lceil \rho L \rceil)}$ (blue triangles) and $\hat{p}_{out}^{(\lceil \rho L \rceil)}$ (red circles). Results are for} $\Delta=0.1$ {(i.e., $p_{{in,out}}=\rho\pm0.05$}) with (a) $L=4$ and (b) $L=64$ layers. We also show effective probabilities for the AND (gold curves) and OR (cyan curves) networks. (Solid and dashed curves give {$\hat{p}_{in}^{(\tilde{L})}$ and $\hat{p}_{out}^{(\tilde{L})}$}, respectively.) {Note for the larger $L$ value in (b) that $\hat{p}_{in}^{(\lceil \rho L \rceil)}$ and $\hat{p}_{out}^{(\lceil \rho L \rceil)}$ have separated and aligned with the asymptotic probabilities $\hat{p}_{in}^{(asym)}$ (blue solid curve) and $\hat{p}_{out}^{(asym)}$ (red dashed curve), respectively}. 
(c)--(f) Adjacency matrices of thresholded networks with $\rho =0.3$, $\Delta=0.1$, $\tilde{L}=\lceil \rho L \rceil$ and various $L$. 
}
\label{fig:spys}
\end{figure}

{In Fig.~\ref{fig:spys}, we illustrate the limiting $L\to\infty$ behavior for thresholded networks with $\tilde{L}=\lceil \rho L\rceil$}. In panels (a)--(b), we plot $\hat{p}_{in}^{(\lceil \rho L\rceil)}$ (blue triangles) and $\hat{p}_{out}^{(\lceil \rho L\rceil)}$ (red circles) versus $\rho$ for $\Delta=0.1$ with (a) $L=4$ and (b) $L=64$. We also plot the effective probabilities {$\hat{p}_{in}^{(\tilde{L})}$} (solid curves) and {$\hat{p}_{out}^{(\tilde{L})}$} (dashed curves) for the AND (gold curves) and OR (cyan curves) networks. In panel (b), we additionally plot the limiting effective probabilities $\hat{p}_{in}^{(asym)}$ (blue solid curve) and $\hat{p}_{out}^{(asym)}$ (red dashed curve). 
{Comparing panel (b) to (a), one can observe that as $L$ increases, the piecewise-continuous solutions $\hat{p}_{{in,out}}^{(\lceil \rho L\rceil)}$ separate and align with the respective asymptotic solutions $\hat{p}_{{in,out}}^{(asym)}$.}

In Figs.~\ref{fig:spys}(c)--(f), we illustrate adjacency matrices {${\bf \hat{A}}^{(\lceil \rho L\rceil)}$} of thresholded networks with $\rho =0.3$ and $\Delta=0.1$ for various $L$. We note that the community structure is undetectable for $L=1$ since $\Delta^*=0.1095$, whereas it is detectable (and visually apparent) for $L=128$.
Comparing (c)--(f) illustrates the $L\to\infty$ limiting behavior of ${\bf \hat{A}}^{(\lceil \rho L\rceil)}$.
{Specifically, application of Hoeffding's inequality \cite{Hoeffding} (and using that $p_{in,out}-\rho = \pm\Delta/2$) yields ${p}_{in}^{(\lceil \rho L\rceil)}   \ge 1 - e^{  -L\Delta^2/2} $ and ${p}_{out}^{(\lceil \rho L\rceil)} \le  e^{ -L\Delta^2/2}$, {which implies that} $\hat{p}_{in}^{(\lceil \rho L\rceil)} \to1$ and $\hat{p}_{out}^{(\lceil \rho L\rceil)} \to0$ with increasing $L$ so that $\hat{A}^{(\lceil \rho L \rceil)}_{ij}\to\delta_{c_ic_j}$, where $\delta_{nm}$ is the Kronecker delta function.}

\begin{figure}[t]
\centering
\epsfig{file = 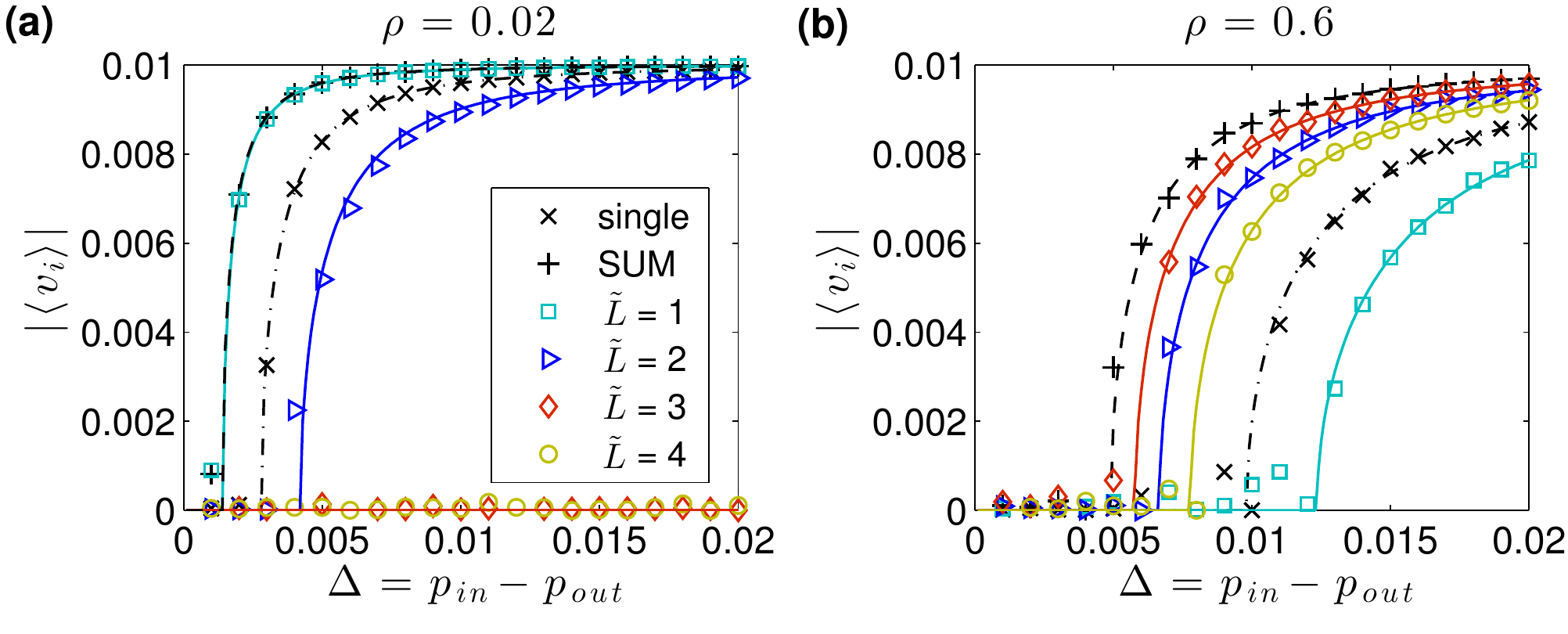, clip =,width=1\linewidth } 

\caption{(Color online) 
{\it Phase transition at $\Delta^*$ for the dominant eigenvector $\bf v$ of the modularity matrix.}
We show observed (symbols) and predicted values given by Eq.~\eqref{eq:evec} (curves) for the mean eigenvector entry {$|\langle v_i \rangle|$} within a community for $N=10^4$ and $L=4$. 
} 
\label{fig:PhaseTransition}
\end{figure}

We conclude by studying the dominant eigenvector $\bf v$ of the appropriate modularity matrix, which undergoes a phase transition at $\Delta^*$: $\{v_i\}$ and the community labels $\{c_i\}$ are uncorrelated for $\Delta<\Delta^*$, whereas they are correlated for $\Delta>\Delta^*$. 
Using methodology developed in \cite{Benaych_2011}, we find that the entries $\{v_i\}$ within a community are Gaussian distributed with mean
%We extend the analysis of \drt{\cite{Nadakuditi_2012}} to multilayer (and potentially dense) 
\begin{equation}
|\langle{v_i}\rangle| = \sqrt{\frac{1}{N}}\sqrt{1-\frac{\lambda_2^2}{(NL\Delta)^2} } ,\label{eq:evec}
\end{equation}
which we use as an order parameter to observe the phase transition. In Fig.~\ref{fig:PhaseTransition}, we depict observed (symbols) and predicted values given by Eq.~\eqref{eq:evec} (curves) of $|\langle{v_i}\rangle|$ for a single layer ($\times$-symbols), the summation network ($+$-symbols) and thresholded networks (open symbols). We focus on a range of $\Delta$ that contains $\Delta^*$ for most aggregation methods. Note for the thresholded networks that there is no simple ordering to $\Delta^*$, which can be deduced by examining Fig.~\ref{fig:Multiple_thresholds}(a) for $\rho\in\{0.02,0.6\}$. Finally, we note that finite-size effects amplify disagreement between observed and predicted values near the phase transitions.

In this Letter, we studied the limitations on community detection for multilayer networks with layers drawn from a common SBM. As an illustrative model, we analyzed the effect of layer aggregation on the detectability limit $\Delta^*$ for two equal-sized communities. When layers are aggregated by summation, we analytically showed detectability is always enhanced and $\Delta^*$ vanishes as $\mathcal{O}(L^{-1/2})$. When layers are aggregated by thresholding this summation, $\Delta^*$ depends sensitively on the choice of threshold, $\tilde{L}$. For $\tilde{L}=\lceil \rho L\rceil$, we analytically found $\Delta^*$ to also vanish as $\mathcal{O}(L^{-1/2})$.
We note that our analysis also describes layer aggregation by taking the mean, $L^{-1}\sum_l{\bf A}^{(l)}$, since the multiplication of a matrix by a constant (e.g., $L^{-1}$) simply scales all eigenvalues by that constant. Thus, our results are in excellent agreement with previous work \cite{Han2015} that proved spectral clustering via the mean adjacency matrix to be a consistent estimator for the community labels. 
%Importantly, averaging yields edge weights that approximate the edge probabilities, and this information can improve community detection algorithms \cite{Han2015,Martin2015}.

Finally, it is commonplace to threshold {pairwise-interaction} data to construct network representations that are sparse and unweighted and can be studied at a lower computational cost. Our research provides insight into this common---yet not well understood---practice.  It is important to extend our work to more-complicated settings. We believe fruitful directions should include allowing the SBMs of layers to be correlated \cite{Abbe2014} (that is, rather than identical) as well as allowing layers to be organized into ``strata'' \cite{Stanley2015}, so that layers within a stratum follow a similar SBM but the SBMs can greatly differ between strata. We are currently extending our analysis to hierarchical SBMs using methodology developed in \cite{Peixoto2013}.

\acknowledgements{The authors were supported by the NIH (R01HD075712, T32GM067553, T32CA201159), the James S. McDonnell Foundation (\#220020315) {and the UNC Lineberger Comprehensive Cancer Center with funding provided by the University Cancer Research Fund via the State of North Carolina.} The content does not necessarily represent the views of the funding agencies. We thank the reviewers for their helpful comments.
}

%%%%%%%%%%%%%%%%%%%%%%%%%%%%%%%%%%%%%%%%%%%%%%%%
%%%%%%%%%%%%%%%%%%%%%%%%%%%%%%%%%%%%%%%%%%%%%%%%
\bibliographystyle{plain}

\pagebreak
\clearpage

%%%%%%%%%%%%%%%%%%%%%%%%%%%%%%%%%%%%%%%%%%%%%%%%%%%%%%
%%%%%%%%%%%%%%%%%%%%%%%%%%%%%%%%%%%%%%%%%%%%%%%%%%%%%%
%
%            supplemental material
%
%%%%%%%%%%%%%%%%%%%%%%%%%%%%%%%%%%%%%%%%%%%%%%%%%%%%%%
%%%%%%%%%%%%%%%%%%%%%%%%%%%%%%%%%%%%%%%%%%%%%%%%%%%%%%

\onecolumngrid
\appendix

\subsection{{\large Supplemental Material: Enhanced detectability of community structure in {multilayer} networks through layer aggregation}}
{\centering

Dane Taylor,$^{1,*}$ Saray Shai,$^1$ Natalie Stanley$^{1,2}$ Peter J. Mucha$^1$ \\
\vspace{.1cm}
{\small \emph{$^1$Carolina Center for Interdisciplinary Applied Mathematics, \\
Department of Mathematics, University of North Carolina, Chapel Hill, NC 27599, USA\\
$^2$Curriculum in Bioinformatics and Computational Biology, \\
University of North Carolina, Chapel Hill, NC 27599, USA}}

\vspace{.5cm}
}

\twocolumngrid
\appendix

%\author{Dane Taylor}
%\email{dane.r.taylor@gmail.com}
%\affiliation{Carolina Center for Interdisciplinary Applied Mathematics, Department of Mathematics, University of North Carolina, Chapel Hill, NC 27599, USA}
%
%\author{Saray Shai}
%\affiliation{Carolina Center for Interdisciplinary Applied Mathematics, Department of Mathematics, University of North Carolina, Chapel Hill, NC 27599, USA}
%
%\author{Natalie Stanley}
%\affiliation{Carolina Center for Interdisciplinary Applied Mathematics, Department of Mathematics, University of North Carolina, Chapel Hill, NC 27599, USA}
%\affiliation{Curriculum in Bioinformatics and Computational Biology, University of North Carolina, Chapel Hill, NC 27599, USA}
%
%\author{Peter J. Mucha}
%\affiliation{Carolina Center for Interdisciplinary Applied Mathematics, Department of Mathematics, University of North Carolina, Chapel Hill, NC 27599, USA}
%
%\title{Supplemental Material: Enhanced detectability of community structure in {multilayer} networks through layer aggregation}
%
%\maketitle

\subsection{Eigenspectra of Modularity Matrix $\overline{B}$}
\vspace{-.3cm}

\begin{figure}[t]
\centering
~\\
~\\
\epsfig{file = 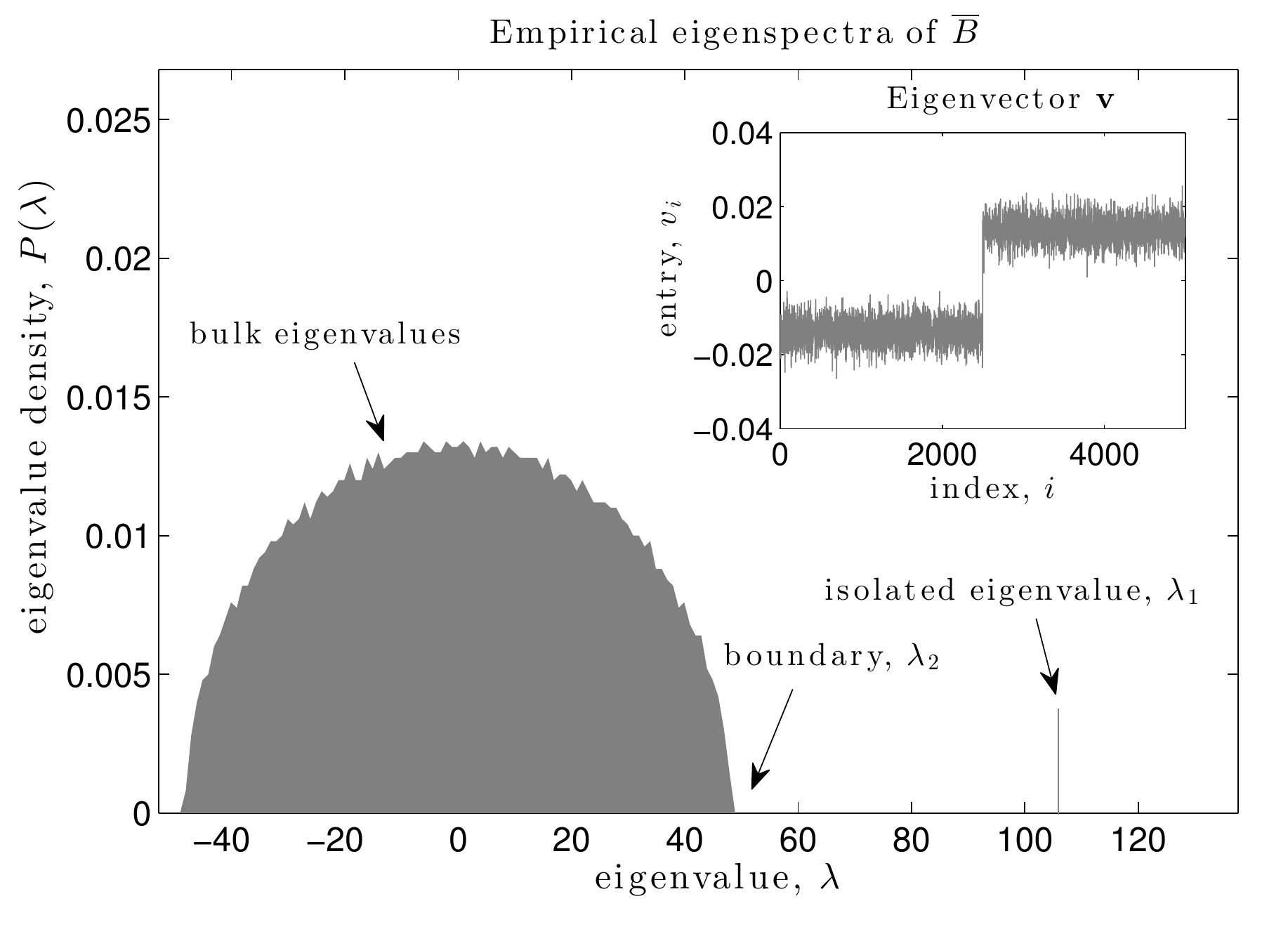, clip =,width=1\linewidth } 
\caption{(Color online) 
{\it Empirical eigenspectra of the modularity matrix ${\bf\overline{B}}$.}
We plot the distribution of eigenvalues of ${\bf\overline{B}}={\bf \overline{A}}-\rho L {\bf 1}{\bf 1}^T$, which consists of two parts: bulk eigenvalues that solve Eq.~\eqref{eq:bulk} and an isolated eigenvalue that solves Eq.~\eqref{eq:isolated}. The subplot depicts the eigenvector $\bf v$ corresponding to the largest eigenvalue $\lambda_1$, which encodes community structure and gives the spectral bi-partition. Results are shown for $N=5000$ nodes, $L=4$ layers, mean edge probability $\rho=0.03$, and probability difference $\Delta=0.01$ (see main text).
 }
\label{fig:evals}
\end{figure}

Here, we provide further details about the limiting $N\to\infty$ distribution of eigenvalues for modularity matrix ${\bf \overline{B}}={\bf \overline{A}}-\rho L {\bf 1}{\bf 1}^T$, where ${\bf 1}$ is a vector of ones, ${\bf \overline{A}}=\sum_l {\bf A}^{(l)}$ is the summation of the layers' adjacency matrices, and each ${\bf A}^{(l)}$ is drawn from a single stochastic block model with two equal-sized communities. Our analysis is based on methodology developed in [1,2],
%\cite{Benaych_2011,Nadakuditi_2013}, 
which we extend to layer-aggregated multiplex networks including those that are potentially dense. As shown in Fig.~\ref{fig:evals}, the spectrum consists of two parts---an isolated eigenvalue $\lambda_1$ (whose corresponding eigenvector ${\bf v}$ encodes the spectral bi-partition) and bulk eigenvalues which have an $N\to\infty$ limiting distribution $P(\lambda)$. 
In the analysis to follow, we will assume that the community structure is detectable. We begin by defining random matrix
\begin{align}
{\bf X} =   {\bf \overline{B}} - \langle {\bf \overline{B}} \rangle ,
\label{eq:X}
\end{align}
where $ \langle \overline{B}_{ij} \rangle $ indicates the mean value of $\overline{B}_{ij}$ across the random matrix ensemble. The decomposition of ${\bf \overline{B}}$ facilitates the analysis of spectra through the following relation,
\begin{align}
0&=\text{det}\left(  z {\bf I} - {\bf \overline{B}}\right) \nonumber\\
 &= \text{det}\left(  z {\bf I} - ( {\bf X} + \langle {\bf \overline{B}} \rangle)\right) \nonumber\\
 &= \text{det}\left(  z {\bf I} - {\bf X} \right) \text{det}\left(  {\bf I} - (z {\bf I}- {\bf X} )^{-1} \langle {\bf \overline{B}} \rangle \right) ,
 \label{eq:two_parts}
\end{align}
which assumes the invertibility of  $ (z {\bf I}- {\bf X} )$.
Equation \eqref{eq:two_parts} highlights that the spectra of ${\bf \overline{B}}$ can be studied in two parts: a distribution $P(z)$ of bulk eigenvalues that solve the first term,
\begin{align}
0= \text{det}\left(  z {\bf I} - {\bf X} \right),\label{eq:bulk}
\end{align} 
and an isolated eigenvalue that solves the second term, 
\begin{align}
0=\text{det}\left(  {\bf I} - (z {\bf I}- {\bf X} )^{-1} \langle {\bf \overline{B}} \rangle \right).\label{eq:isolated}
\end{align}

Before describing the solutions to Eq.~\eqref{eq:bulk} and Eq.~\eqref{eq:isolated}, we comment on the matrices ${\bf X}$ and $\langle {\bf\overline{B}} \rangle$. Recall that each entry $\overline{A}_{ij}$ follows a binomial distribution [see Eq.~(2) in the main text], so that their mean and variance is  
\begin{align}
\langle  {A}_{ij}  \rangle  &=  \left\{\begin{array}{cc} Lp_{in}  ,&c_i=c_j \\ Lp_{out} ,&c_i\not=c_j .\end{array}\right.  \nonumber\\
\langle  {A}_{ij}^2\rangle -\langle  {A}_{ij}\rangle^2  &= \left\{\begin{array}{cc} Lp_{in}(1-p_{in}) ,&c_i=c_j \\ Lp_{out}(1-p_{out}),&c_i\not=c_j ,\end{array}\right. 
\label{eq:X2}
\end{align}
where $c_i,c_j\in\{1,2\}$ indicate the community labels of nodes $i$ and $j$. It follows that $\{{X}_{ij}\}$ have mean and variance  
\begin{align}
\langle {X}_{ij}  \rangle  &= 0 \nonumber\\
\langle {X}_{ij}^2\rangle &= \left\{\begin{array}{cc} Lp_{in}(1-p_{in}) ,&c_i=c_j \\ Lp_{out}(1-p_{out}),&c_i\not=c_j .\end{array}\right. 
\label{eq:X2}
\end{align}
We next consider $\langle {\bf\overline{B}}\rangle$. Using that $\overline{B}_{ij} = \overline{A}_{ij}-\rho L$ and $\rho = (p_{in}+p_{out})/2$ (i.e., $p_{in,out}-\rho = \pm\Delta/2$), we find 
\begin{align}
\langle \overline{B}_{ij}\rangle  &= \left\{\begin{array}{cc}  L\Delta/2,&c_i=c_j \\  -L\Delta/2 ,&c_i\not=c_j .\end{array}\right. 
\label{eq:B1}
\end{align}
Importantly, $\langle {\bf\overline{B}} \rangle$ is a rank-one matrix [2]%\cite{Nadakuditi_2013}
\begin{align}
\langle \overline{B} \rangle  = \theta_1 {\bf u}{\bf u}^T ,%- Lp_{in} I,
\label{eq:B_define}
\end{align}
where ${\bf u} =N^{-1/2} [1,\dots,1,-1,\dots,-1]^T$ and
\begin{align}
\theta_1=\frac{NL\Delta }{2}. \label{eq:theta} 
\end{align}
We point out that without loss of generality, we have assumed that nodes $\{1,\dots,N/2\}$ are in community 1 (i.e., $u_i>1$ for these nodes) and nodes $\{1+N/2,\dots,N\}$ are in community 2 (i.e., $u_i<1$ for these nodes).

We now return our attention to solving Eq.~\eqref{eq:two_parts} for the eigenvalues of ${\bf\overline{B}}$. We first solve Eq.~\eqref{eq:bulk} to study the bulk eigenvalues. The limiting $N\to\infty$ spectral density $P(z)$ of ${\bf X}$ can be solved via its average resolvent $\langle (z {\bf I} - {\bf X})^{-1} \rangle$ and the Stieltjes transform [1]%\cite{Benaych_2011}
\begin{align}
P(z) = \frac{-1}{N\pi} \text{Im} ~\text{Tr} \langle (z {\bf I} - {\bf X})^{-1} \rangle ,\label{eq:stieljes}
\end{align}
where $z \in\mathbb{C}$ approaches the real line $\mathbb{R}$ from above. Our analysis of Eq.~\eqref{eq:stieljes} directly follows the methodology presented in [2],
%\cite{Nadakuditi_2013}, 
albeit for an aggregated multiplex network and allowing for potentially dense networks. In particular, the average resolvent can be expanded as
\begin{align}
\text{Tr} \langle (z {\bf I} - {\bf X})^{-1} \rangle = \frac{1}{z}  \sum_{k=0}^\infty \frac{\text{Tr}\langle {\bf X}^k \rangle} {z^k},
\end{align}
where 
\begin{align}
\text{Tr} \langle {\bf X}^k \rangle &= \sum_{i_1,\dots,i_k } \langle {X}_{i_1i_2}{X}_{i_2i_3}\dots {X}_{i_ki_1} \rangle,\label{eq:ah1}
\end{align}
and the sequence $\{i_1,i_2,\dots,i_k,i_1\}$ defines an Euler tour at node $i_1$.
Because $\langle {X}_{ij} \rangle=0$, any term in Eq.~\eqref{eq:ah1} that contains a variable just once will be mean zero across the ensemble. Moreover, terms containing a variable more than twice become negligible when the nodes' degrees are large. As shown in [2],
%\cite{Nadakuditi_2013}, 
the only terms remaining are those that contain each variable exactly twice and for which $k$ is even, implying that
\begin{align}
\text{Tr} \langle (z {\bf I} - {\bf X})^{-1} \rangle %= \frac{1}{z} = \sum_{k=0}^\infty \frac{\text{Tr}\langle X^k \rangle} {z^k}\nonumber\\
 &\approxeq  \frac{1}{z}  \sum_{k=0}^\infty \frac{\text{Tr}\langle {\bf X}^{2k} \rangle} {z^{2k}},\label{eq:cata}
\end{align}
where 
 \begin{align}
\text{Tr} \langle {\bf X}^{2k} \rangle &\approxeq \sum_{i_1,\dots,i_k } \langle{X}_{i_1i_2}^2 {X}_{i_2i_3}^2 \dots  {X}_{i_ki_1}^2 \rangle \nonumber\\
&= N\left({NL\tilde{p}}\right)^{k} C_k, \label{eq:CAT} %.\label{eq:final_trace}
\end{align}
$C_k$ is the Catalan number, and
\begin{align}
\tilde{p}&= [p_{in}(1-p_{in})+p_{out}(1-p_{out})]/2\nonumber\\
&=\rho(1-\rho) -  {\Delta^2}/{4}\label{eq:ptilde}
\end{align}
is the average variance across the matrix entries $\{X_{ij}\}$. We note for sparse networks that $\tilde{p} \approx \rho$, which was the case considered by [2]. After substituting Eq.~\eqref{eq:CAT} into Eq.~\eqref{eq:cata}, we obtain $\text{Tr} \langle (z {\bf I} - {\bf X})^{-1} \rangle  =  t(z) $, where
\begin{align}
t(z) = \frac{z - \sqrt{z^2 - \lambda_2^2} }{\lambda_2^2/2},\label{eq:t}
\end{align}
and 
\begin{align}
\lambda_2=\sqrt{4NL\tilde{p}}, \label{eq:lam}
\end{align}
which recovers Eq.~(4) in the main text. Moreover, we substitute Eq.~\eqref{eq:t} into Eq.~\eqref{eq:stieljes} to obtain Eq.~(3) in the main text.

We now study the isolated eigenvalue $\lambda_1$ by solving the ensemble average of Eq.~\eqref{eq:isolated}, 
\begin{align}
0= \text{det}\left(  I - \langle (z {\bf I}- {\bf X} )^{-1}\rangle  \langle {\bf \overline{B}}  \rangle \right)  ,
\label{eq:second_term}
\end{align}
Because $\langle (z{\bf I}- {\bf X} )^{-1}\rangle_{ij}=0$ for $i\not=j$, we have that $\langle (z{\bf I}- {\bf X} )^{-1}\rangle = t(z){\bf I}$.
It follows that $t(z) \theta_1$ is the largest eigenvalue of $\langle (z {\bf I}- {\bf X} )^{-1}\rangle  \langle {\bf \overline{B}}  \rangle$. Because Eq.~\eqref{eq:second_term} requires this matrix to have an eigenvalue equal to one, we find that $\lambda_1$ solves
\begin{align}
1 = t(\lambda_1) \theta_1.\label{eqq}
\end{align}
Using that $t(z)$ has the inverse $t^{-1}(z) =  {z}^{-1}+ {z\lambda_2^2}/{4}$, we solve Eq.~\eqref{eqq} for $\lambda_1$ to obtain
\begin{align}
\lambda_1 &= t^{-1}(\theta_1^{-1})  \nonumber\\
&= {\theta_1} + \frac{\lambda_2^2}{4\theta_1}  \label{eq:Max_eval}.
\end{align}
After substituting the definition of $\theta_1$ given by Eq.~\eqref{eq:theta}, we recover Eq.~(5) in the main text. 
Setting $\lambda_1=\lambda_2$ gives the solution $\lambda_2=2\theta_1$, which recovers Eq.~(6) in the main text.
As shown in [1],
%\cite{Benaych_2011}, 
the corresponding eigenvector ${\bf v}$ is correlated with ${\bf u}$, which can be measured by the inner product
\begin{align}
|{\bf u}^T\bf{v}|^2& = 1 - \frac{\lambda_2^2}{4\theta_1^2}. \label{eq:innerprod}
\end{align}
We note that in the large $N$ limit,
\begin{align}
\frac{|{\bf u}^T{\bf v}|}{N^{1/2}} 
&\approx \left|\frac{1}{N/2}\sum_{i=1}^{N/2} v_i   \right| ,
\end{align}
where the right hand side is the mean entry within a community.
Therefore, we divide Eq.~\eqref{eq:innerprod} by $N$ and take the square root to obtain Eq.~(9) in the main text.

\vspace{.5cm}
{\small 
\noindent [1] F. Benaych-Georges and R. R. Nadakuditi, Adv. in Math. {\bf227}, 494 (2011).
\vspace{-.00cm}

\noindent [2] R. R. Nadakuditi and M. E. J. Newman, Phys. Rev. Lett. {\bf108}(18), 188701 (2012).
}

%\bibliographystyle{plain}
%\begin{thebibliography}{99}
%\bibitem{Benaych_2011} F. Benaych-Georges and R. R. Nadakuditi, {Adv. in Math.} {\bf227}, 494 (2011).
%\bibitem{Nadakuditi_2013} R. R. Nadakuditi and M. E. J. Newman, {Phys. Rev. Lett.} {\bf108}(18), 188701 (2012).
%\end{thebibliography}
%
%\clearpage
\end{document}